\documentclass[prb,onecolumn,aps,preprint,superscriptaddress,showpacs,floatfix]{revtex4-2}
\usepackage{amsbsy,amssymb,amsmath,bm}
\usepackage{graphicx,color,epsfig,rotate}
\usepackage{xspace,units}
\usepackage{textcomp}		
\usepackage{epstopdf}
\usepackage{ulem}
\usepackage{tabularx}

\usepackage{multirow}		
\usepackage{dcolumn}							
\newcolumntype{d}[1]{D{.}{.}{#1}}
 
\begin{document}
\title{In situ straining of epitaxial freestanding ferroic films by a microelectromechanical device}

\author{Simone Finizio}
\email{simone.finizio@psi.ch}
\affiliation{Paul Scherrer Institut, 5232 Villigen PSI, Switzerland}

\author{Tim A. Butcher}
\email{tim.butcher@mbi-berlin.de}
\affiliation{Paul Scherrer Institut, 5232 Villigen PSI, Switzerland}
\affiliation{Max Born Institute for Nonlinear Optics and Short Pulse Spectroscopy, 12489 Berlin, Germany}

\author{Maria Cocconcelli}
\affiliation{Dipartimento di Fisica, Politecnico di Milano, Via G. Colombo 81, 20133 Milano, Italy}

\author{Elisabeth M\"uller}
\affiliation{Paul Scherrer Institut, 5232 Villigen PSI, Switzerland}

\author{Lauren J. Riddiford}
\affiliation{Paul Scherrer Institut, 5232 Villigen PSI, Switzerland}
\affiliation{Laboratory for Mesoscopic Systems, Department of Materials, ETH Zurich, 8093 Zurich, Switzerland}

\author{Jeffrey A. Brock}
\affiliation{Paul Scherrer Institut, 5232 Villigen PSI, Switzerland}
\affiliation{Laboratory for Mesoscopic Systems, Department of Materials, ETH Zurich, 8093 Zurich, Switzerland}

\author{Chia-Chun Wei}
\affiliation{Department of Physics, National Cheng Kung University, Tainan 70101, Taiwan}

\author{Li-Shu Wang}
\affiliation{Department of Physics, National Cheng Kung University, Tainan 70101, Taiwan}

\author{Jan-Chi Yang}
\affiliation{Department of Physics, National Cheng Kung University, Tainan 70101, Taiwan}
\affiliation{Center for Quantum Frontiers of Research \& Technology (QFort), National Cheng Kung University, Tainan 70101, Taiwan}

\author{Shih-Wen Huang}
\affiliation{Paul Scherrer Institut, 5232 Villigen PSI, Switzerland}

\author{Federico Maspero}
\affiliation{Dipartimento di Fisica, Politecnico di Milano, Via G. Colombo 81, 20133 Milano, Italy}

\author{Riccardo Bertacco}
\affiliation{Dipartimento di Fisica, Politecnico di Milano, Via G. Colombo 81, 20133 Milano, Italy}

\author{J\"org Raabe}
\affiliation{Paul Scherrer Institut, 5232 Villigen PSI, Switzerland}

\date{\today}

\begin{abstract}

\noindent 
Mechanical strain can be used to control physical properties in materials. The experimental investigation of strain-induced effects at the nanoscale is of importance not only for its fundamental aspects, but also for the development of device applications. Transmission X-ray microscopy is a particularly well-suited technique for nanoscale imaging of magnetic materials, but its compatibility with \textit{in-situ} mechanical straining of samples is limited. In this work, we present a setup for applying tailored \textit{in-situ} mechanical strains to freestanding thin films by means of a micro electromechanical system (MEMS) actuator. We then present a proof-of-concept experiment in which a freestanding 80 nm thick (001) BiFeO$_3$ multiferroic thin film is strained with the MEMS device, allowing us to control the coupled ferroelectric/spin cycloidal configuration.
\end{abstract}

\maketitle

\section{Introduction}

In many classes of materials, strain can couple into several ferroic orders, allowing one to control it by deforming the material, or to strain the material by controlling the ferroic order through its conjugate field \cite{Spaldin2005, Eerenstein2006}. Examples of these classes of materials include piezoelectric (strain couples to the ferroelectric order) and magnetoelastic (strain couples to the magnetic order) systems, both of which not only have widespread everyday applications, but are also of interest in fundamental research, making the understanding of the behavior of the coupling between strain and other ferroic orders at the nanoscale of great interest.

Examples of magnetoelastic effects, the investigation of which requires a microscopy technique include the study of domain wall and vortex dynamics as a function of applied strain \cite{Finizio2017, Foerster2017, Filianina2019}, the study of strain-induced phase transitions in antiferromagnetic materials \cite{Jani2024, Harrison2025}, and the investigation of the effect of strain on the coupling mechanisms in ferroic systems \cite{Haykal2020, Zhao2006, Hu2016, Meyer2024, Wang2025, Segantini2025}. Their investigation requires the combination of a non-invasive microscopy technique able to resolve nanometric features with contrast mechanisms that allow for the visualization of the desired ferroic order. Thanks to the possibility of combining non-invasive imaging with strong dichroic contrast mechanisms (X-ray magnetic circular or X-ray linear dichroism - XMCD/XLD), with the possibility of attaining nanometric resolutions, and with the option to apply \textit{in-situ} excitations, soft X-ray microscopy techniques such as photoemission electron microscopy (PEEM) \cite{Buzzi2013, Finizio2014, Filianina2019}, scanning transmission X-ray microscopy (STXM) \cite{Finizio2017, Jani2024, Harrison2025}, and X-ray ptychography \cite{Butcher2024, Neethirajan2024} are an ideal match. In this work, we will concentrate on STXM and ptychography.


STXM and X-ray ptychography are transmission microscopy techniques. Due to the low penetration depth of soft X-rays, only sample thicknesses under one micrometer can be investigated. Using phase contrast imaging at photon energies below the elemental absorption edges of the sample, the thickness that can still be imaged with dichroic contrast can be extended to the micrometer range \cite{Neethirajan2024}, but not significantly beyond that. Therefore, the standard method employed for straining samples in non-transmission microscopy techniques, i.e. by means of a piezoelectric crystal actuator \cite{Finizio2014, Buzzi2013, Foerster2017}, cannot be used for transmission X-ray microscopy due to the non-negligible thickness of the piezoelectric crystal. 


In recent years, a solution for \textit{in-situ} mechanical straining of X-ray transparent samples was developed by modifying a gas cell used for atmospheric chemistry investigations \cite{Huthwelker2011}. With this method, a 50 nm thick Si$_3$N$_4$ membrane is bent by pressurizing the gas cell, which buckles towards the region with lower pressure, generating a tensile strain at the top surface, and a compressive strain at the bottom surface, which is then transferred to structures on the membrane \cite{Finizio2016, Harrison2025}. Due to the fabrication constraints, the typical geometries of Si$_3$N$_4$ membranes are square or rectangular, i.e. the strain is applied primarily along the two axes of the square/rectangle \cite{Finizio2016, Finizio2017, Harrison2025}. By tailoring the membrane geometry, it is possible e.g. to simulate an effective uniaxial strain application \cite{Finizio2016, Harrison2025}, or to simulate a situation as close as possible to a circular membrane, resulting in the generation of an isotropic strain \cite{Finizio2016}. 



Membrane bending has been used for the uniaxial straining of Co$_{40}$Fe$_{40}$B$_{20}$ microstructured elements stabilizing a Landau flux-closure state \cite{Finizio2016}. Using a rectangular membrane, a uniaxial magnetic anisotropy due to the magnetoelastic effect could be generated \cite{Finizio2016, Finizio2017}. Another example is the demonstration of strain-driven control of the antiferromagnetic configuration in freestanding $\alpha$-Fe$_2$O$_3$ thin films \cite{Jani2024, Harrison2025}. 


While STXM investigations combined with bending of Si$_3$N$_4$ membranes were possible \cite{Finizio2017, Harrison2025}, this setup has limitations that reduce its applicability to epitaxial systems. The main limitation is that the Si$_3$N$_4$ membranes are amorphous and can therefore not be used as substrates for the growth of epitaxial films. While single-crystalline Si membranes are commercially available and two dimensional nanosheets used as growth template layer can allow for the growth of some epitaxial material \cite{Le2020, Nguyen2016}, this still holds several experimental challenges. In addition, the Si$_3$N$_4$ membrane geometries are restrictive: while certain complex geometries such as e.g. circular membranes are feasible \cite{Serra2016}, it is not possible to fabricate membranes with arbitrary geometries, limiting the ensemble of strain configurations that can be generated.

As no commercial solutions for the \textit{in-situ} straining of thin membranes in conditions compatible with soft X-ray microscopy exist, in this work we present a different approach to the \textit{in-situ} straining of X-ray transparent samples, based on the use of micro electromechanical system (MEMS) devices. A lamella was attached to these by means of ion-beam assisted deposition of carbon layers in a focused ion beam (FIB) tool, fixing the two ends of the lamella on the MEMS device. By electrically poling the MEMS device, the X-ray transparent lamella can be strained to values significantly above those achievable either by piezoelectric crystal actuators or by membrane bending. Below, we present the device fabrication and the proof-of-concept experiment demonstrating the application of a tensile strain onto a freestanding BiFeO$_3$ (BFO) lamella, leading to strain-induced motion of ferroelectric domain walls and changes to the coupled spin cycloid state.

\section{In-situ straining with a MEMS device}

\begin{figure*}[!hbt]
    \includegraphics[width=0.9\textwidth]{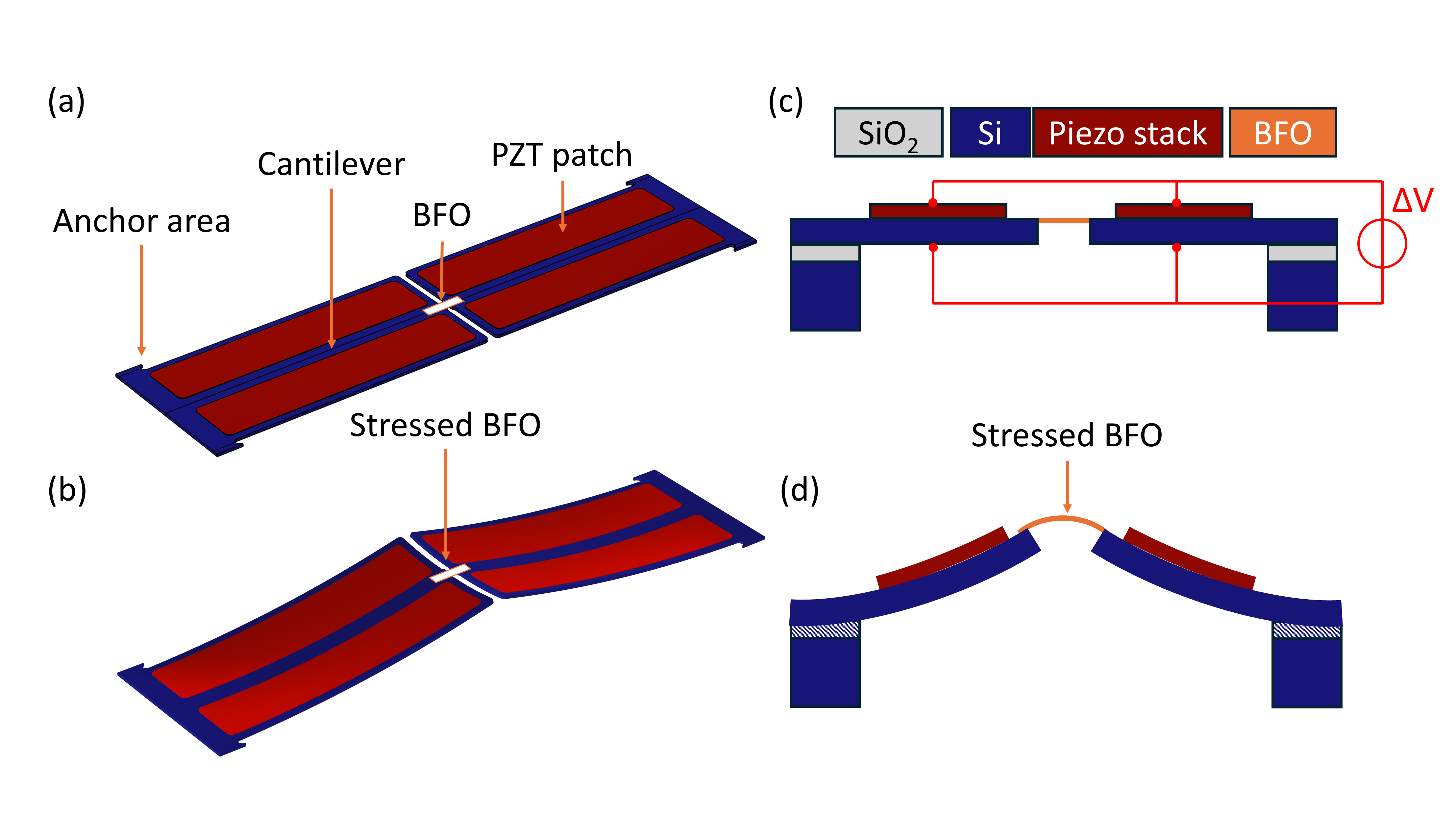}
    \caption{Schematic representation of the MEMS strainer (a) before and (b) after applying a voltage; (c-d) Cross-sectional view of the same device highlighting the different materials. The positions of the contacts used for the voltage poling are sketched by the red lines in (c).}
    \label{fig:MEMS}
\end{figure*}

A sketch of the MEMS devices used in this work is shown in Fig. \ref{fig:MEMS}. The devices were fabricated using the P$\mathrm{\epsilon}$Tra Thin-Film-Piezoelectric technology developed by STMicroelectronics \cite{Ghisi2021}. The actuator consists of a pair of MEMS cantilevers connected by a central bridge. Each cantilever is made of 10 \textmu m thick Si and presents two patches of 2 \textmu m thick Pb(Zr$_{0.52}$Ti$_{0.48}$)O$_3$ (PZT) for a total of four patches connected in parallel. Upon applying a voltage across its thickness, the PZT layer contracts causing the bilayer to deflect upwards. The applied voltage is unipolar to avoid depolarization of the PZT and ensure higher linearity \cite{Opreni2021}.

In order to perform the soft X-ray ptychography imaging experiments, the BFO lamellas had to be positioned on a X-ray transparent region of the sample. The original design of the MEMS devices described above does not have one of such regions, as the two cantilevers are connected through a small 10 \textmu m thick bridge. This bridge was therefore removed by means of plasma focused ion beam (PFIB) cutting, using a 500 nA/30 kV Xe beam in a ThermoFisher Helios 5 Laser Hydra UX instrument hosted at the ScopeM laboratories of ETH Z\"urich, resulting in a 15-25 \textmu m cut separating the two cantilevers. As the cantilever is electrically contacted from both sides, its functionality was unaffected by the PFIB cut.


For the proof-of-concept experiments described in this work, a lamella extracted from a 80 nm thick freestanding (001) BFO film needed to be transferred onto the gap between the two cantilevers. Epitaxial BFO/Sr$_3$Al$_2$O$_6$ (SAO) heterostructures were grown on (001) oriented SrTiO$_3$ (STO) substrates using a pulsed laser deposition system (Mobile Combi-Laser MBE MC-LMBE, Pascal Co., Ltd.). A KrF excimer laser ($\lambda$ = 248 nm) with an energy of 200 mJ and a repetition rate of 10 Hz was utilized for all growth processes. First, a 25 nm sacrificial layer of SAO was deposited on the STO substrate at 740 °C under an oxygen pressure of 45 mTorr. Subsequently, an 80 nm BFO layer was deposited on top of the SAO layer at 650 °C under an oxygen pressure of 65 mTorr. Following the growth, the sample was annealed at 400 $^\circ$C in nearly 1 atm of oxygen for 20 minutes to reduce oxygen vacancies, followed by a slow cooling process to room temperature under the same oxygen environment. Finally, the sample was immersed in deionized water to dissolve the sacrificial SAO layer, allowing the freestanding BFO film to be released and transferred from the solution onto a TEM grid with a lacy carbon support for subsequent XLD analysis.

\begin{figure}[!hbt]
    \includegraphics[width=0.3\textwidth]{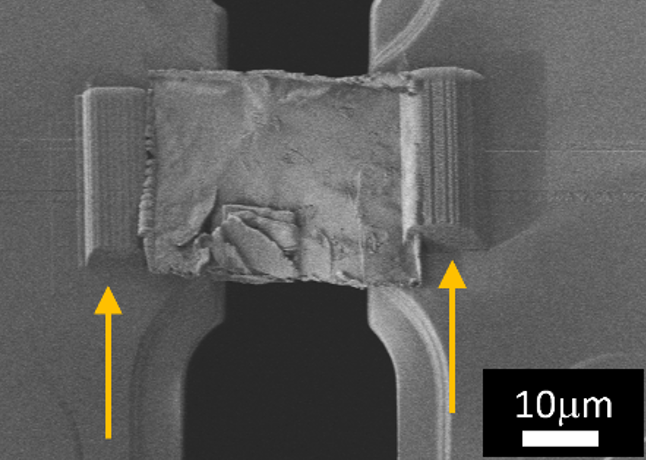}
    \caption{SEM image of a 80 nm thick (001) BFO lamella bridging the gap between the two cantilevers on the MEMS device. The BFO lamella was attached by ion beam assisted carbon deposition in the two points marked by the orange arrows.}
    \label{fig:BFO_SEM}
\end{figure}

For the transfer of the BFO freestanding films, a Ga-FIB instrument (Zeiss NVision40) equipped with a micromanipulator from Kleindieck was employed. As the freestanding films are deposited onto a standard 3 mm diameter Cu mesh after their release from the substrate in a potentially random in-plane orientation, it is necessary to first determine the orientation of the film, in order to guarantee that the lamella is cut along the desired [1-10]/[110] directions. To achieve this, the freestanding film was measured in advance by means of XLD ptychography to identify the crystallographic directions of the BFO film from the orientation of the spin cycloid and from the orientation of the edge of the film \cite{Butcher2024,Butcher2025b}. During the extraction of the lamella, minimal damage from the ion beam has been achieved by using a 10 pA/30 kV beam current for all steps where the membrane had to be imaged with the ion beam. Two line cuts at a distance of 10-20 \textmu m and a length of 80-100 \textmu m were performed with a focused Ga-ion beam set to a 300 pA/30 kV beam current. The micromanipulator needle was then attached with ion-assisted carbon deposition close to the middle of the lamella, which helped to reduce the bending of the very thin stripe after cutting its end off from the rest of the membrane. After cutting the lamella free and lifting it from the TEM-grid, it was fixed with ion-assisted deposition of carbon on both sides of the gap on the MEMS device, followed by cutting off the manipulator needle. The attachment of the manipulator to the middle of the stripe had the additional advantage of helping to not damage the MEMS cantilever structures upon cutting off the manipulator needle from the stripe. A scanning electron micrograph (SEM) of the transferred lamella is shown in Fig. \ref{fig:BFO_SEM}.

The imaging of the evolution of the ferroelectric domains and the associated spin cycloidal configuration in the BFO lamella as a function of the applied strain was also performed with XLD ptychography \cite{Butcher2024}. The measurements were carried out at the SoftiMAX beamline of the MAX IV Laboratory, using a ptychography endstation optimized for the acquisition of high-resolution images \cite{Butcher2025}. The imaging was performed by tuning the X-ray radiation to the second of the crystal-field split L$_3$ peaks at the Fe L-edge, which corresponds to the e$_\textrm{g}$ state of the hybridized O 2p- Fe 3d states \cite{Butcher2025b} (L$_\mathrm{3b}$ edge, ca. 707.5 eV), acquiring two images with the electric field of the linearly-polarized X-ray beam oriented at 0 and 90 degrees in the surface plane, and calculating the asymmetry ratio between the two images (calculated as $(I_0-I_{90})/(I_0+I_{90})$, with $I_0$ and $I_{90}$ being the images acquired at 0 and 90 degree electric field orientations). 

For the imaging experiments described here, a $514 \times 514$ px$^2$ Eiger-iLGAD (inverse low-gain avalanche diode) soft X-ray 2D detector with a square pixel size of 75 \textmu m was utilized \cite{Baruffaldi2025}. The sample-detector distance was selected to be 9.6 cm and the samples were scanned following a Fermat spiral \cite{Huang2014} pattern with a step size of 50 nm, with a selected square region of interest (ROI) of $1.5 \times 1.5$ \textmu m$^2$ size. Diffraction patterns were acquired at each step of the Fermat spiral with an exposure time of 200 ms, using a beam spot of ca. 400 nm diameter. The positioning of the sample was guaranteed through a piezoelectric stage with an interferometric feedback \cite{Holler2015, Butcher2025}.

\begin{figure}[!hbt]
    \includegraphics[width=0.45\textwidth]{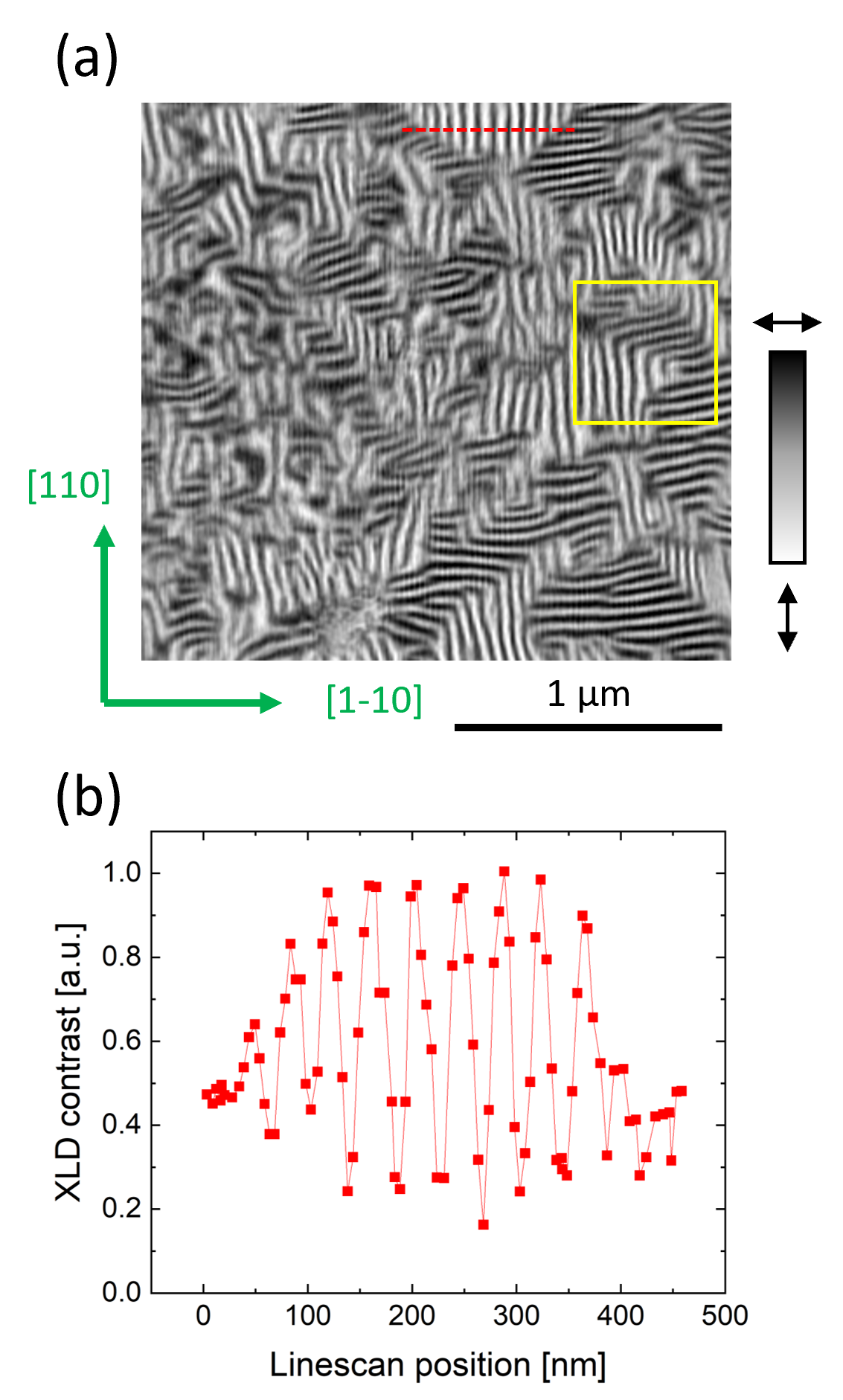}
    \caption{(a) Amplitude contrast XLD-ptychography image of a 80 nm (001) BFO freestanding film. Both the ferroelectric mosaic domains (larger domains) and the spin cycloid (smaller ``waves”) can be observed. The grayscale bar depicts the direction of the ferroelectric polarization. The ROI marked in yellow corresponds to the region shown in Fig. \ref{fig:cycloid}. (b) Linescan across the region marked by the red dotted line in (a), showing the change in XLD contrast caused by the spin cycloid (oscillating part) and by the coupled ferroelectric domain (constant offset from 0).}
    \label{fig:BFO_ptycho}
\end{figure}

The images were then reconstructed from the recorded diffraction patterns using the PtychoShelves software package \cite{Wakonig2020} with the combined difference map \cite{Thibault2008} and maximum likelihood algorithms \cite{odstrcil_2018}. In order to account for the incomplete coherence of the X-ray beam, three probe modes were utilized \cite{thibault_2013}. The ptychographic reconstruction yielded a complex-valued image, of which the amplitude contrast was selected for the ptychographic XLD images presented here. An example of a reconstructed image of the BFO lamella is shown in Fig. \ref{fig:BFO_ptycho}(a), with the XLD contrast variations caused by the spin cycloid shown in Fig. \ref{fig:BFO_ptycho}(b).

The strain-dependent experiments were performed by poling the MEMS device using a Keithley 2450 sourcemeter to supply the voltage. Voltage differences between 0 and 20 V were applied in 2.5 V steps. For each step, the ferroelectric and spin cycloidal configuration of several ROIs on the BFO lamella were imaged. A first estimation of the applied strain magnitude was performed by measuring the displacement of the two cantilevers. Note that the BFO lamella was damaged in this stage of the experiments due to the application of an excessive strain at a voltage difference of ca. 17.5-20 V (corresponding, as shown below in Fig. \ref{fig:MEMS_SEM} to an applied tensile strain of about 2\%).


In order to determine the exact magnitude of the uniaxial tensile strain generated by the MEMS device, the MEMS devices were installed in a Zeiss Supra 55VP SEM and imaged as a function of the applied voltage. The voltage was applied by installing an electrical feedthrough flange on the SEM and connecting the two contacts of the MEMS device to a Keithley 2400 sourcemeter installed outside of the SEM vacuum chamber. The measurements described below were performed either in \textit{post-mortem} conditions, i.e. after the transferred BFO lamella was broken during the X-ray imaging experiments, or in virgin conditions, i.e. without a lamella bridging the gap between the two cantilevers. Therefore, these measurements depict the maximum applicable strain on a given voltage applied to the MEMS device.

\begin{figure}[!hbt]
    \includegraphics[width=0.4\textwidth]{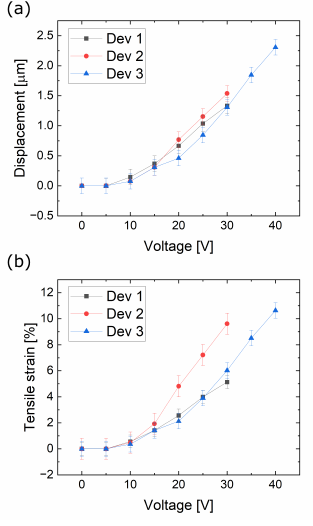}
    \caption{(a) Variation of the distance between the two MEMS cantilevers as a function of the applied voltage difference between the top and bottom surfaces of the cantilever determined from SEM images as a function of the applied voltage. The recorded displacement is independent of the size of the gap between the two cantilevers; (b) Maximum tensile strain generated by the displacement of the cantilevers.}
    \label{fig:MEMS_SEM}
\end{figure}

The MEMS devices were poled with voltages from 0 to 40 V in steps of 5 V. For each voltage step, an SEM image of each cantilever was acquired (images in the Supplementary Information \cite{suppInfo}). The voltage-dependent displacement of the cantilevers was determined from the images and is shown in Fig. \ref{fig:MEMS_SEM}(a). Three different cantilevers, marked as devices 1--3 in the figure, with two differently sized gaps, were investigated. As can be observed in Fig. \ref{fig:MEMS_SEM}(a), the displacement does not depend on the size of the cut, but only on the applied voltage. Note that the errors in the gap distance are due to the pixel size in the SEM images. The size of the cut determines the maximum strain that can be achieved, as shown in Fig. \ref{fig:MEMS_SEM}(b), where the applied tensile strain is defined as $\Delta x/x_0$, being $\Delta x$ the displacement, and $x_0$ the gap size at no applied voltage. As device 2 has a smaller gap (about 15 \textmu m, against about 25 \textmu m for the other two devices), the achievable strain is higher than for the other two cantilever devices. All of the MEMS devices shown in Fig. \ref{fig:MEMS_SEM} can generate significantly larger tensile strains than both single-crystal piezoelectric substrates and membrane bending geometries and comparable to the current state-of-the-art for perovskite films, which is about 8\% \cite{Hong2020}.

The tensile strain generated by the MEMS device follows a parabolic dependence with respect to the applied voltage (see Fig. \ref{fig:MEMS_SEM}). If the same device is used for hundreds of cycles, the strain-vs-voltage response changes with respect to the first cycles. This change, typically driven by wake-up/conditioning effects linked to the reorganization of the ferroelectric domain walls and defects in the piezoelectric material of the MEMS, lead to a linearization of the strain-vs-voltage response, which needs to be taken into consideration if the same sample is investigated in several measurement campaigns \cite{Opreni2021, Nguyen2021}.

\section{Rotation of ferroelectric domain walls by tensile strain}

The image shown in Fig. \ref{fig:BFO_ptycho} was acquired in the as-transferred configuration, i.e. without any applied strain. Upon poling the MEMS device, a tensile strain of varying magnitudes was applied to the BFO lamella. At each strain level, gauged as described above from the displacement of the cantilevers, an XLD-ptychography image of the ferroelectric and spin cycloidal configuration of the BFO lamella was acquired. While no global changes in either the orientation or the period of the spin cycloid of the BFO lamella were observed, several regions existed where the mechanical straining led to ferroelectric domain wall motion.

\begin{figure}[!hbt]
    \includegraphics[width=0.45\textwidth]{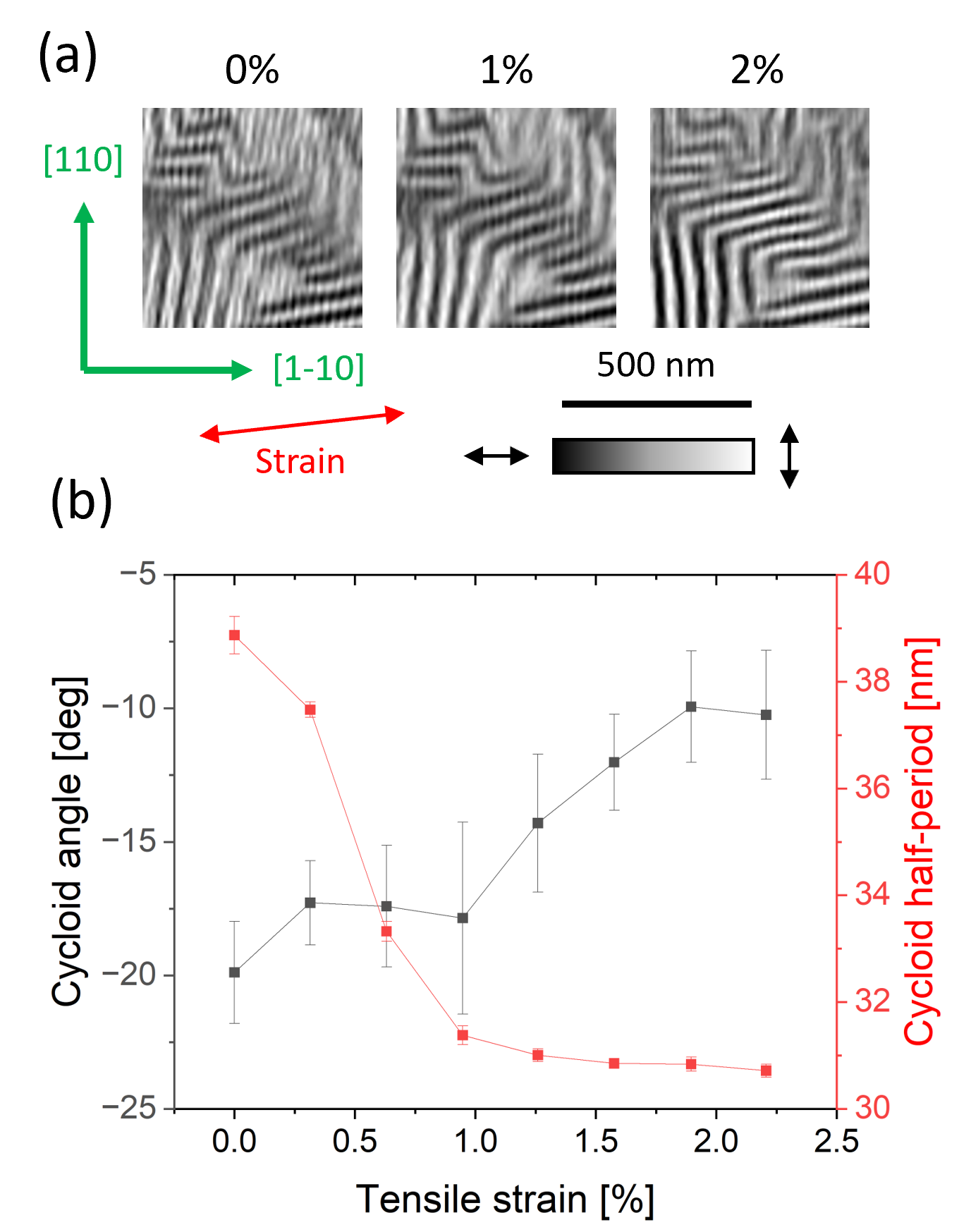}
    \caption{(a) XLD ptychography images as a function of strain applied to a 80 nm (001) BFO lamella in a ROI where a ferroelectric domain wall motion was observed, marked by the yellow square in Fig. \ref{fig:BFO_ptycho}(a). Associated with the domain wall motion, a change in both the period and propagation direction of the spin cycloidal state can be observed. The grayscale bar depicts the direction of the ferroelectric polarization, while the red arrow marks the direction of the applied tensile strain, oriented at about at 10$^\circ$ from the [1-10] direction); (b) Half-period and orientation of the cycloid in (a) as a function of the applied strain. Note that the BFO lamella ruptured at an applied strain of ca. 2\%.}
    \label{fig:cycloid}
\end{figure}

Fig. \ref{fig:cycloid} shows one of the regions where ferroelectric domain wall motion was observed. For this domain wall, the motion occurred upon the application of a tensile strain of about 1\%, as shown in Fig. \ref{fig:cycloid}(a). The two ferroelectric domains separated by the moving domain wall exhibit both a rotation of the spin cycloid propagation vector and a reduction of the cycloid period, likely indicating a rotation of the coupled ferroelectric polarization direction either towards the strain axis or perpendicular to it. Similar changes occur for other ferroelectric domains in the imaged ROI (see Supplementary Information \cite{suppInfoBFO}). A change in the cycloid period was also observed for substrate-bound BFO films when increasing the tensile strain component if compared to the growth conditions on a SrTiO$_3$ substrate \cite{Haykal2020}, indicating that similarities in the behavior of the spin cycloid with increasing tensile strain can be found between substrate-bound and freestanding films.

Due to the rupturing of the BFO lamella at an applied tensile strain of ca. 2\%, which was likely preceded by permanent plastic deformation, a full verification of the reversibility and reproducibility of the ferroelectric domain wall motion and associated spin cycloid changes could not be carried out. Nonetheless, it was still possible to provide a first proof-of-concept of the functionality of the MEMS device described above for the application of large \textit{in-situ} tensile strains in a soft X-ray ptychography imaging experiment.


\section{Conclusion and Outlooks}

In conclusion, we have shown a design for a MEMS device that can be used for the soft X-ray ptychographic imaging of \textit{in-situ} strained epitaxial freestanding films. The concept is based on a dual-cantilever MEMS device with a 15--25 \textmu m gap between the cantilevers, where an epitaxial film lamella is positioned by means of a FIB instrument. By poling the MEMS device, a tensile strain in the axis parallel to the cantilevers can be transferred to the lamella, whilst preserving the X-ray transparency requirement for ptychography imaging. Tensile strains reaching several percentage points were obtained, i.e. above those achievable with other methods compatible with transmission microscopy. A proof-of-concept measurement was performed with a 80 nm thick (001) BFO lamella, where the application of a tensile strain of ca. 1\% led to the combined motion of several ferroelectric domain walls, also causing a change in the orientation and period of the spin cycloid coupled with the ferroelectric order in the BFO.

\begin{figure}[!hbt]
    \includegraphics[width=0.45\textwidth]{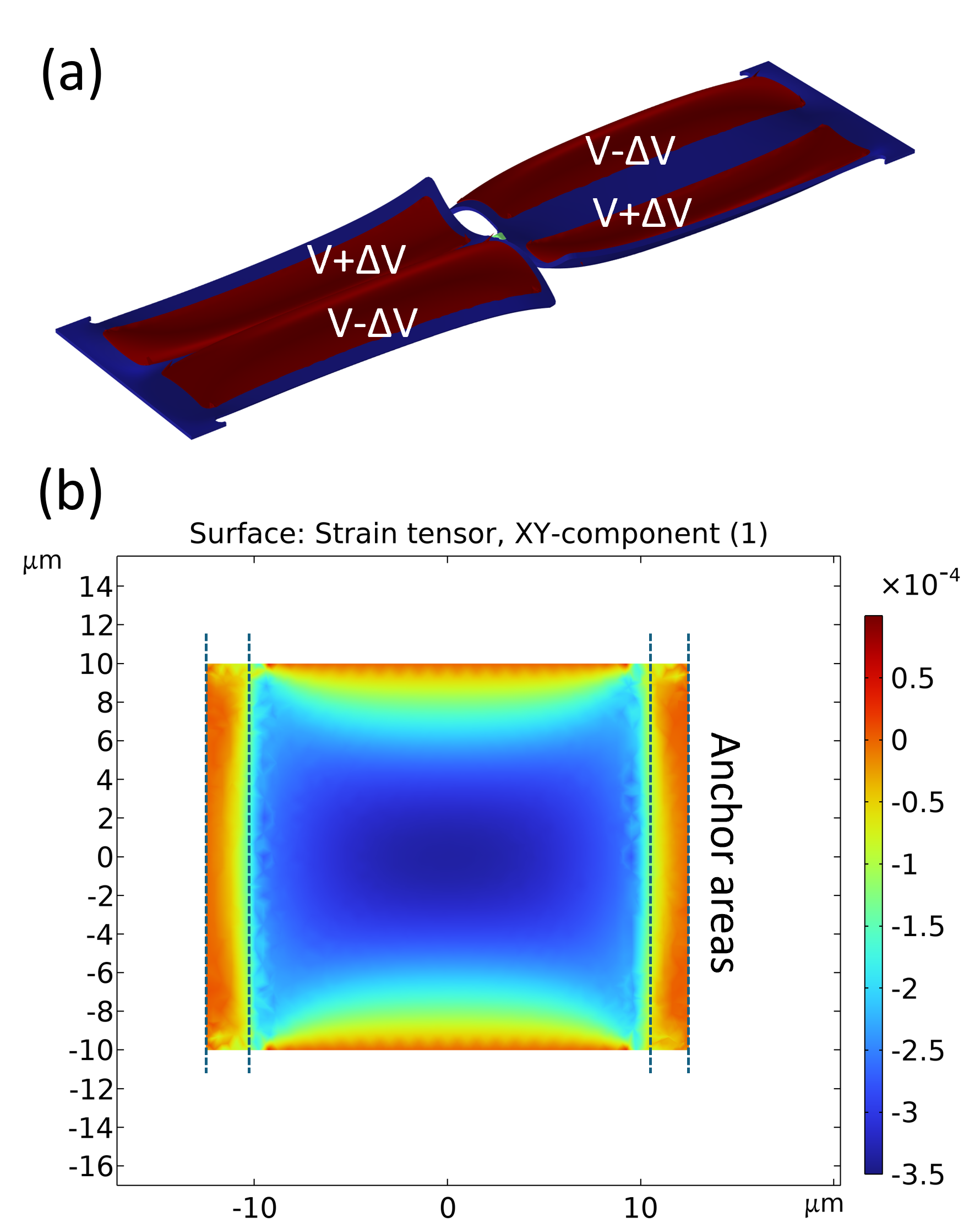}
    \caption{(a) Possible scheme for a shear stress actuation; (b) Shear stress map on a film positioned between the two cantilevers.}
    \label{fig:shear}
\end{figure}

The design shown above allows one to apply a tensile strain to lamellas. However, other types of strain such as compressive or shear strain are also of interest. To generate shear strain, the same MEMS device geometry could be exploited by actuating the piezoelectric patches with differential voltage as sketched in Fig. \ref{fig:shear}. This could induce a rotation of the two cantilevers and a subsequent torque on the film under test. Additionally, a direct gauging of the applied strain could be implemented by measuring the displacement of the cantilever with the piezoresistive sensor integrated with the actuator, as shown in \cite{Frigerio2021, Shin2010, Dukic2015}. This would allow one not only to directly gauge the applied strain \textit{in-situ}, but also to significantly reduce the error in the determined strain if compared to \textit{post-mortem} SEM investigations. Another possible improvement is in the use of a piezoelectric material such as AlN, which could ensure a more linear response of the system, thus reducing the need for an \textit{in-situ} gauging of the strain by means of focus or linescans whilst also enabling bipolar actuation.

Finally, another possible advantage of using MEMS actuators for the \textit{in-situ} straining of materials is that their mechanical resonances can be controlled: while the resonance frequency for the device shown in this work is relatively low (the first bending mode is around 10 kHz), MEMS devices can be designed with resonance frequencies up to the MHz range \cite{Bagolini2023}, allowing us to explore dynamical processes at MHz frequencies, such as e.g., the dynamics of the magnetoelastic effect \cite{Foerster2017}. Such designs, if modified to be compatible with transmission microscopy, would allow us to access the MHz frequency range and perform time-resolved investigation of magnetoelastic processes, both with standard STXM time-resolved imaging \cite{Puzic2010, Finizio2020} and, if the resonance frequency is tuned to one of the harmonics of the master clock of the synchrotron lightsource, time-resolved ptychography \cite{Butcher2025a}. 

\section*{Acknowledgments}

Soft X-ray ptychography measurements were performed with the SOPHIE endstation at the SoftiMAX beamline of the MAX IV Laboratory. Development of SOPHIE was supported by the Swiss Nanoscience Institute (SNI). Research conducted at MAX IV, a Swedish national user facility, is supported by the Swedish Research council under contract 2018-07152, the Swedish Governmental Agency for Innovation Systems under contract 2018-04969, and Formas under contract 2019-02496. T.A.B. acknowledges funding from SNI and the European Regional Development Fund. J.-C.Y. acknowledges the financial support from National Science and Technology Council (NSTC) in Taiwan, under grant no. NSTC 112-2112-M-006-020-MY3. L.R. acknowledges funding from the ETH Zurich Postdoctoral Fellowship Program 22-2 FEL-006. The authors also thank MSSORPS Co., Ltd. for high-quality TEM sample preparation and preliminary examination. The research is also supported in part by Higher Education Sprout Project, Center for Quantum Frontiers of Research \& Technology (QFort) at National Cheng Kung University, Taiwan. We thank I. Beinik, K. Th\r{a}nell, and J. Schwenke for support in the operation of the SoftiMAX beamline at Max IV. We thank E. Fr\"ojdh, F. Baruffaldi, M. Carulla, J. Zhang and A. Bergamaschi for the development of the iLGAD Eiger detector. The iLGAD sensors were fabricated at Fondazione Bruno Kessler (Trento, Italy). The authors acknowledge Joakim Reuteler of the ScopeM laboratory at ETH Z\"urich for his assistance with the PFIB lithography of the MEMS devices. The authors acknowledge STMicroelectronics for providing the MEMS actuator. We acknowledge the usage of instrumentation provided by the Electron Microscopy Facility (EMF) at PSI and thank the EMF team for their help and support.

\providecommand{\noopsort}[1]{}\providecommand{\singleletter}[1]{#1}%

\end{document}